\documentclass{elsart} 
\usepackage{graphicx}
\usepackage{amsmath}
\usepackage{amssymb}
\usepackage{graphicx}

\begin{document}

\begin{frontmatter}

\title{Dynamics of a quantum measurement}
\author{Armen E. Allahverdyan$^{1,3)}$}
\author{\underline{Roger Balian\thanksref{thank1}}$^{2)}$} 
and
\author{Theo M. Nieuwenhuizen$^{1)}$}
\address{$^{1)}$ Institute for Theoretical Physics,
Valckenierstraat 65, 1018 XE Amsterdam}
\address{$^{2)}$
SPhT,
CEA-Saclay,
F-91191 Gif sur Yvette Cedex, France}

\address{$^{3)}$ Yerevan Physics Institute,
Alikhanian Brothers St. 2, Yerevan 375036, Armenia}
\thanks[thank1]{
Corresponding author. 
E-mail: balian@spht.saclay.cea.fr}

\begin{abstract}
We work out an exactly solvable hamiltonian model which retains all
the features of realistic quantum measurements. In order to use an
interaction process involving a system and an apparatus as a
measurement, it is necessary that the apparatus is macroscopic. This
implies to treat it with quantum statistical mechanics. The relevant
time scales of the process are exhibited. It begins with a very rapid
disappearance of the off-diagonal blocks of the overall density matrix
of the tested system and the apparatus. Possible recurrences are
hindered by the large size of the latter. On a much larger time scale
the apparatus registers the outcome:
Correlations are established between the final values of the pointer
and the initial diagonal blocks of the density matrix of the tested
system. We thus derive Born's rule and von Neumann's reduction of the
state from the dynamical process.
\end{abstract}

\begin{keyword}
quantum measurements \sep quantum dynamics
\PACS 05.30.-d \sep 05.70.Ln
\end{keyword}
\end{frontmatter}

%\twocolumn

\section{Properties of quantum measurement processes.}

The interpretation of quantum mechanics is tightly connected with the
understanding of quantum measurements. The textbooks and most articles
devoted to this subject focus on the initial and final states of the
system $\mathrm{S}$ and the apparatus $\mathrm{A}$, without describing
in detail the coupled dynamics of $\mathrm{S}$ and $\mathrm{A}$ during
the measurement process \cite{wh,balian}. Our purpose is to study such
a dynamics. We shall present a model which displays all the features
of a realistic quantum measurement and is exactly solvable
\cite{ABN}. This will allow us to explain how these features arise
from the microscopic equations of motion. We have therefore to solve a
problem of quantum statistical mechanics for which it is crucial to
take into account both the microscopic nature of the object
$\mathrm{S}$ and the macroscopic nature of the apparatus $\mathrm{A}$.

A quantum measurement shares with classical measurements several
general properties. It is an \textit{experiment} during which the
system $\mathrm{S}$ to be tested and the apparatus $\mathrm{A}$,
separately prepared at an initial time $t_{\rm i}=0$,
\textit{interact}. This dynamical process creates correlations between
the state of $\mathrm{A}$ at the final time $t_{\mathrm{f}}$ and the
initial state of $\mathrm{S}$. Information can thereby be gained on
$\mathrm{S}$ through observation, at (or after) the time
$t_{\mathrm{f}}$, of some pointer variable belonging to
$\mathrm{A}$. Provided the value of the pointer variable is
\textit{registered}, the ${\rm r\hat{o}le}$ of the observer is
eliminated.

Quantum measurements differ from classical ones in two respects. On
the one hand, quantum mechanics is an irreducibly
\textit{probabilistic} theory. What is called \textquotedblleft state
of a system\textquotedblright, whether it is represented by a
wavefunction, a ket or a density operator, refers to a
\textit{statistical ensemble} of systems, all prepared under the same
conditions as the system in hand. 
This object gathers our whole information about the
preparation of any system belonging to this ensemble. However, in contrast to
a state in classical physics, a quantum state necessarily involves statistical
fluctuations due to the non-commutation of the observables and is therefore
akin to a probability distribution. Indeed, if two physical quantities are
represented by observables $A$ and $B$ with commutator $\left[  A,B\right]
=2iC$, they must statistically fluctuate in agreement with Heisenberg's
inequality $\Delta A\Delta B\geq\left\vert \left\langle C\right\rangle
\right\vert $ if the expectation value $\left\langle C\right\rangle $ is
non-zero in the considered state. Accordingly, a quantum measurement
must in general involve statistical fluctuations. In classical
physics, probabilities may occur in measurements due to uncertainties
in the preparation of the initial state and to measurement errors;
however one can imagine more and more precise preparations, and more
and more precise measurements, so that nothing forbids to find the
outcome within a negligible error. In a quantum measurement, even
under the most perfect conditions, the outcome is always probabilistic.
Consider a measurement of the observable 
$\hat{s}=\sum_is_i\hat{\Pi}_i$ of
$\mathrm{S}$; we denote as $s_{i}$ its eigenvalues, as $\hat{\Pi}_{i}$
its eigenprojections in the Hilbert space of $\mathrm{S}$, and as
$\hat{A}$ the pointer observable of the apparatus $\mathrm{A}$ which
is coupled to $\hat{s}$. 
(For a non-degenerate eigenvalue of $\hat{s}$ with eigenfunction
$|\psi_i\rangle$, $\hat{\Pi}_i=|\psi_i\rangle\langle\psi_i|$.)
If the system is prepared initially in an
eigenstate of $\hat{s}$ corresponding to the eigenvalue $s_{i}$, the
pointer variable takes the well defined value $A_{i}$. However, for an
arbitrary initial state of $\mathrm{S}$ represented at the initial
time by the density operator $r\left( 0\right) $, different runs of
the experiment, performed on systems identically prepared in this state
$r\left( 0\right) $, may yield different outcomes $A_{i}$. The
probability $p_{i}$ of finding $A_{i}$ at the time $t_{\mathrm{f}}$,
for the statistical ensemble described by $r\left( 0\right) $, is
given by \textit{Born's rule}%
\begin{equation}
p_{i}=\operatorname{Tr}_{\rm S}\,\hat{\Pi}_{i}r\left(  0\right)  \text{ ,}%
\tag{1.1}\label{001.001}%
\end{equation}
which exhibits the irreducibly probabilistic nature of the
measurement.  (The situation is not different for a pure state
$r(0)=|\phi\rangle\langle\phi|$, in which case $p_i=
\langle\phi|\hat{\Pi}_i|\phi\rangle$.)

On the other hand, the \textit{perturbation} of $\mathrm{S}$ induced by a
measurement cannot be neglected in quantum physics, whereas  
nothing prevents in classical physics to make it smaller and smaller.
Consider first the apparatus. If we include in $\mathrm{A}$ the registration
device, any measurement, whether classical or quantal, must perturb
$\mathrm{A}$ so as to be informative. This perturbation, when induced by a
microscopic system $\mathrm{S}$, should be sufficiently strong so as to let
$\mathrm{A}$ undergo a macroscopic change. In many real measurements, this
interaction process drastically modifies the system $\mathrm{S}$ itself and
may even destroy it. We wish to focus on \textit{ideal measurements}, those
which perturb $\mathrm{S}$ as little as possible. This is achieved by
preparing $\mathrm{A}$ at the initial time in a metastable state, with density
operator $\mathcal{R}\left(  0\right)  $. During the measurement, the
interaction with $\mathrm{S}$ triggers $\mathrm{A}$ in the same way as a small
source, leading it towards one or another among several possible stable states
$\mathcal{R}_{i}$, each characterized by a value $A_{i}$ of the pointer
variable. 
The fact that these possible final states are exclusive is expressed
by ${\rm Tr}\,\mathcal{R}_{i}\mathcal{R}_{j}=\delta_{ij}$,
which for positive matrices $\mathcal{R}_{i}$ and $\mathcal{R}_{j}$
implies 
\begin{equation}
\mathcal{R}_{i}
\mathcal{R}_{j}=0,\quad {\rm for}\quad i\not =j.
\nonumber
\end{equation}
Such a property holds, in particular, if A is a large system with
spontaneously broken invariance, $\mathcal{R}_{i}$ being then an
equilibrium state characterized by the value ${\rm Tr}
(\hat{A}\,\mathcal{R}_{i})=A_i$ of its order parameter.

In an ideal measurement, the occurrence of $A_{i}$ is correlated
with the fact that $\mathrm{S}$ lies  
at the time $t_{\mathrm{f}}$ in an eigenstate of $\hat{s}$ associated with
the eigenvalue $s_{i}$. More precisely, \textit{von Neumann's reduction}
expresses that the initial joint density operator $\mathcal{D}\left(
0\right)  =r\left(  0\right)  \otimes\mathcal{R}\left(  0\right)  $ of the
compound system $\mathrm{S}+\mathrm{A}$ is transformed at the final time
$t_{\mathrm{f}}$ into%
\begin{gather}
\mathcal{D}\left(  0\right) \mapsto\mathcal{D}\left(  t_{\mathrm{f}}\right)
%\nn\\ 
=\sum_{i}\left[  \hat{\Pi}_{i}r\left(  0\right)  \hat{\Pi}_{i}\right]
\otimes\mathcal{R}_{i}\text{ .}\tag{1.2}\label{001.002}%
\end{gather}
Full memory is kept of the diagonal blocks $\hat{\Pi}_{i}r\left(  0\right)
\hat{\Pi}_{i}$ of the initial state $r\left(  0\right)  $ of
$\mathrm{S}$. This means that the statistics of the {\it any}
observable which {\it commutes} with $\hat{s}$ is left unchanged by
the measurement process. However, 
the initial information carried 
by the off-diagonal blocks of 
$r(0)$ is lost in spite of
the ideal nature of the measurement.

The expression (1.2) encompasses (1.1), which is obtained by taking
the expectation value of $\hat{\Pi}_{i}$ over the state
$\mathcal{D}\left( t_{\mathrm{f}}\right) $. It is more detailed since
it describes the overall properties of the final state, but it applies
only to the ideal measurements for which the perturbation of
$\mathrm{S}$ is the weakest. It exhibits a correlation between the
possible final states of $\mathrm{A}$ and those of $\mathrm{S}$, which
is expressed by
\begin{equation}
\langle \hat{\Pi}_i(\hat{A}-A_i)^2\rangle=0,
\tag{1.3}\label{1.3}
\end{equation} 
meaning that
$\hat{A}$ takes the value $A_i$, when $\hat{s}$ equals $s_i$.

From the form (\ref{001.002}) 
of $\mathcal{D}(t_{\rm f})$
we can infer that the observation of the
pointer variable allows us to split the statistical ensemble initially
described by $\mathcal{D}\left( 0\right) $ into a set of subensembles,
each characterized by a value $A_{i}$ of the pointer variable. In each
subensemble, the state of $\mathrm{S}+\mathrm{A}$ is factorized, with
$\mathrm{S}$ represented by the normalized density operator
$\hat{\Pi}_{i}r\left( 0\right) \hat{\Pi}_{i}/p_{i}$ and $\mathrm{A}$
by $\mathcal{R}_{i}$. Splitting the ensemble into subensembles labeled
by $i$ thus decorrelates $\mathrm{S}$ from $\mathrm{A}$.
\textit{Selecting} one subensemble indexed by $A_{i}$ constitutes a
\textit{preparation} of $\mathrm{S}$ in the projected state $\hat{\Pi}%
_{i}r\left(  0\right)  \hat{\Pi}_{i}/p_{i}$, which will characterize for
future experiments the statistics of the subensemble. A further ideal
measurement of $\hat{s}$ will then leave this state unchanged while providing
a perfect prediction of the outcome $A_{i}$. 

It is therefore crucial to
\textit{loose the information} included in the off-diagonal terms $\hat{\Pi
}_{i}r\left(  0\right)  \hat{\Pi}_{j}$ ($i\neq j$)\ present in the initial
state $r\left(  0\right)  $ to achieve the measurement of $\hat{s}$: This is
a price to pay in order to {\it gain information} about the eigenvalues $s_{i}$ of
the observable $\hat{s}$. It is also crucial to \textit{register} $A_{i}$ so
as to filter the final ensemble into subensembles: This is needed to determine
the probabilities $p_{i}$, proportional to the number of counts of $A_{i}$,
and to prepare $\mathrm{S}$ in an eigenstate of $\hat{s}$ if the measurement
is ideal.

\section{Irreversibility of measurements}

Due to the above two unavoidable features, the transformation from
$\mathcal{D}\left(  0\right)  $ to $\mathcal{D}\left(  t_{\mathrm{f}}\right)
$ is \textit{irreversible}. This property is exhibited, in particular, by the
entropy balance. The von Neumann entropy 
\begin{equation}
S[\mathcal{D}\left(  t_{\mathrm{f}}\right)  ]  =S\left[  \sum _{i}\hat{\Pi
}_{i}r\left(  0\right)  \hat{\Pi}_{i}\right]  +\sum _{i}p_{i}S\left[
\mathcal{R}_{i}\right] \tag{2.1}\label{002.001}%
\end{equation}
of the final state 
\footnote{The expression (\ref{002.001}) is found by using (\ref{001.002}),
which implies (using $\mathcal{R}_{i}\mathcal{R}_{j}=0$ for $i\not =j$)
that the entropy is a sum of contributions arising from each $i$.
Note that if the eigenvalues of $\hat{s}$ are non-degenerate and if
the states $\mathcal{R}_i$ 
of A describe canonical equilibrium with spontaneously broken
invariance, the state (\ref{001.002}) of S+A is the one which
maximizes entropy under the following conditions: (i) the statistics
of $\hat{s}$ is the same as for $r(0)$ (this fixes $p_i$); (ii)
the expectation value of the energy of A is given (this characterizes
the various states $\mathcal{R}_i$ with the same entropy as being canonical
ones); the system and the apparatus are correlated according to 
(\ref{1.3}) (this sets the diagonal elements of the state of S in
correspondence with the $\mathcal{R}_i$'s ).
}
is indeed larger than that%
\begin{equation}
S\left[  \mathcal{D}\left(  0\right)  \right]  =S\left[  r\left(  0\right)
\right]  +S\left[  \mathcal{R}\left(  0\right)  \right] \tag{2.2}%
\label{002.002}%
\end{equation}
of the initial state for two reasons. 
On the one hand, when
the density operator $r\left(
0\right)  $ includes off-diagonal blocks $\hat{\Pi}_{i}r\left(  0\right)
\hat{\Pi}_{j}$ ($i\neq j$), their truncation raises the entropy. 
On the other hand, a
robust registration requires that the possible final states $\mathcal{R}_{i}$
of $\mathrm{A}$ are more stable than the initial state $\mathcal{R}\left(
0\right)  $, so that their entropy is larger. The latter effect dominates
because the apparatus is necessarily macroscopic, as discussed below. 

Another manifestation of the irreversibility of the process is the
fact that two different initial states of $\mathrm{S}$ which have the
same diagonal blocks lead to the same final state $\mathcal{D}\left(
t_{\mathrm{f}}\right) $.

We aim at explaining how the state of $\mathrm{S}+\mathrm{A}$ switches from
$\mathcal{D}\left(  0\right)  $ to $\mathcal{D}\left(  t_{\mathrm{f}}\right)
$ during the time lapse $t_{\mathrm{f}}$. We wish to rely \textit{only} on the
basic laws of quantum mechanics, applied to the system $\mathrm{S}+\mathrm{A}%
$. Namely, (i) the physical quantities are represented by hermitian operators
$\hat{X}$ acting in the Hilbert space of $\mathrm{S}+\mathrm{A}$; (ii) a state
of the system $\mathrm{S}+\mathrm{A}$ (or more precisely of the statistical
ensemble to which this system belongs) is represented at each time by a
density operator $\mathcal{D}$, and this state implements the correspondence
between any observable and its expectation value as $\langle \hat
{X}\rangle $ $=\operatorname{Tr}\hat{X}\mathcal{D}$; (iii) since
$\mathrm{S}+\mathrm{A}$ is isolated, its joint density operator $\mathcal{D}$
evolves according to the Liouville -- von Neumann equation%
\begin{equation}
i\hbar\frac{d\mathcal{D}}{dt}=\left[  \hat{H},\mathcal{D}\right]  \text{
,}\tag{2.3}\label{002.003}%
\end{equation}
generated by the Hamiltonian $\hat{H}$. This equation is expected to govern
the dynamical process leading from $\mathcal{D}\left(  0\right)  $ to
$\mathcal{D}\left(  t_{\mathrm{f}}\right)  $, which we wish to analyze. One
cannot hope to solve this question in the general case of real
measurements, and this led many authors to consider models \cite{models}.
Here we work out a model that retains all the features of
realistic measurements, so as to understand how the process can be 
interpreted as a measurement.

We have, however, to cope with the celebrated \textit{measurement problem}.
The equation of motion (2.3) generates for $\mathcal{D}\left(  t\right)  $ a
unitary evolution which conserves von Neumann's entropy $S=-\operatorname{Tr}%
\mathcal{D}\ln\mathcal{D}$ of $\mathrm{S}+\mathrm{A}$, in contradiction with
the consequence (2.1-2) of von Neumann's surmise (1.2).

This problem is akin
to the \textit{paradox of irreversibility} in statistical mechanics, which
relies on the contradiction between the reversibility of the microscopic
evolution of a system and the irreversibility of its macroscopic behaviour.
The solution of this paradox is based as well known on the large number $P$ of
particles of the system. Remember that $\mathcal{D}\left(  t\right)  $
represents at each time our whole statistical information about this system,
including the correlations between any number of particles. We then associate
with $\mathcal{D}$ for given $p<P$ a reduced coarse-grained density operator
$\tilde{\mathcal{D}}^{\left(  p\right)  }$, 
which is equivalent to $\mathcal{D}$ as
regards all the \textquotedblleft simple\textquotedblright\ observables, that
is, all the expectation values of quantities correlating at most $p$ particles,
but which is completely random as regards the higher order correlations
\cite{Mayer}. 
This density operator 
$\tilde{\mathcal{D}}^{\left( p\right) }$ should contain no
more amount of information than the minimum required to account for the
\textquotedblleft simple\textquotedblright variables; its entropy is
therefore larger than that of $\mathcal{D}$ and may increase. Although
$\mathcal{D}$ and $\mathcal{\tilde{D}}^{\left(  p\right)  }$ are
indistinguishable for all practical purposes since high order correlations are
inobservable, their entropies can differ significantly for $P\gg1$ because
these irrelevant correlations are numerous. As the time $t$ flows, the
interactions build up in $\mathcal{D}\left(  t\right)  $ through the evolution
(2.3) correlations between larger and larger numbers of particles. For nearly
all physical systems and models, provided $1\ll p\ll P$, the high order
correlations are so intricate that they do not affect the subsequent evolution
of the \textquotedblleft simple\textquotedblright\ variables, at least over
any reasonable time scale. It is then legitimate to replace $\mathcal{D}
\left(  t\right)  $ by its reduction $\mathcal{\tilde{D}}^{\left(  p\right)
}\left(  t\right)  $, which evolves irreversibly. This irreversibility is
interpreted as a leak of information in $\mathcal{D}\left(  t\right)  $ from
the \textquotedblleft simple\textquotedblright\ variables towards the
correlations of more than $p$ particles. Such correlations are ineffective,
and the order that they carry cannot return in practice to the
\textquotedblleft simple\textquotedblright\ variables, because recurrence
times are inaccessibly large even for a few tens of particles. This argument
can be made mathematically rigorous in solvable models by letting first $P$,
then $p$ tend to infinity \cite{Mayer}.  

We shall propose a similar solution for the measurement problem. It will be
essential to deal with the microscopic system $\mathrm{S}$ exactly. However,
here again, the apparatus $\mathrm{A}$ will be a macroscopic object, with
$P\gg 1$ degrees of freedom. For finite $P $ we shall perform
\textit{approximations}, which amount to the above-mentioned replacement of
$\mathcal{D}$ by $\tilde{\mathcal{D}}^{\left(  p\right)  }$, and which are
\textit{necessary} to account for irreversibility. 
It can be shown \cite{ABN} that the \textit{errors} thus introduced in
the solution of (2.3) \textit{become negligible in the limit}
$P\rightarrow\infty$ as regards all the observables involving a finite
number of degrees of freedom. The expression (1.2) that we shall
derive below for the final state $\mathcal{D}\left(
t_{\mathrm{f}}\right) $ will actually be valid for all components of
$\mathcal{D}$, except for those describing the irrelevant correlations
between an extremely large number of degrees of freedom. Discarding those
imperceptible but numerous correlations is legitimate, although they
are responsible for the difference
between the values of the entropies (2.1) and (2.2).

\section{The model}

In our model, the system $\mathrm{S}$ is the simplest possible: a spin
$\frac{1}{2}$. The obervable $\hat{s}=\hat{s}_{z}$ to be measured is the
$z$-component of this spin, with eigenvalues $s_{\uparrow}=+1$, $s_{\downarrow
}=-1$. In the $i=\uparrow\downarrow$ basis, the initial  
state of $\mathrm{S}$ is represented by the $2\times2$ density matrix
$r\left(  0\right)  $.

The apparatus $\mathrm{A}$ simulates a \textit{magnetic dot}, a collection of
spins in a small solid grain. It can be analyzed into two elements, the
\textit{magnet} $\mathrm{M}$ and the \textit{phonon bath} $\mathrm{B}$. The
magnet $\mathrm{M}$ consists of $N\gg1$ spins with Pauli operators
$\hat{\sigma}_{a}^{\left(  n\right)  }$ ($a=x,y,z$\ ; $n=1,\ldots N$). These
spins interact according to a Curie--Weiss type of Hamiltonian%
\begin{equation}
\hat{H}_{\mathrm{M}}=-\frac{1}{4}JN\hat{m}^{4}\text{ ,}\tag{3.1}%
\label{003.001}%
\end{equation}
where $\hat{m}\equiv\frac{1}{N}\sum_{n}\hat{\sigma}_{z}^{\left(  n\right)  }%
$\ is the magnetization per spin in the $z$-direction. This is a model for
superexchange interactions in metamagnets, suited for a small anisotropic grain.

This part of the apparatus will describe the pointer. The choice of (3.1) as
its Hamiltonian relies on the following remarks. The measurement process wil
be governed by a coupling between the system $\mathrm{S}$ and the apparatus
$\mathrm{A}=\mathrm{M}+\mathrm{B}$, which we represent as the spin-spin
interaction%
\begin{equation}
\hat{H}_{\mathrm{SA}}=-g\hat{s}_{z}
\sum_{n}\hat{\sigma}_{z}^{\left(  n\right)  }
\hat{s}_{z}
=-g\hat{s}_{z}
N\hat{m}\text{ .}\tag{3.2}\label{003.002}%
\end{equation}
Seen from the viewpoint of $\mathrm{M}$, (3.2) looks like the effect of an
operator-valued magnetic field along $z$. 

As a preliminary step, 
let us consider the \textit{equilibrium} of the system $\mathrm{M}$
\textit{alone} at the temperature $T$. If $\mathrm{M}$ is submitted to an
external magnetic field, its Hamiltonian is the sum of (3.1) and (3.2) where
$\hat{s}_{z}$\ is replaced by $s_{i}=+1$ or $-1$\ for $i=\uparrow$\ or
$\downarrow$. In the limit $N\rightarrow\infty$\ the \textit{static
mean-field} approach is exact, a property which suggests that the dynamics of
a model including $\mathrm{S}$ and $\mathrm{A}=\mathrm{M}+\mathrm{B}$ could
also be solved exactly. The equilibrium of $\mathrm{M}$ is found as well known
by looking for the minimum of the free energy per spin%
\begin{equation}
F_{i}\left(  m\right)  =-s_{i}gm-\frac{1}{4}Jm^{4}-TS\left(  m\right)  \text{
,}\tag{3.3}\label{003.003}%
\end{equation}
\begin{gather}
S\left(  m\right)  =-\frac{1+m}{2}\ln\frac{1+m}{2}%\nonumber\\
-\frac{1-m}{2}\ln\frac
{1-m}{2}\text{ ,}\tag{3.4}\label{003.004}%
\end{gather}
expressed in terms of the \textit{order parameter} $m$. The index $i$\ refers
to the direction $s_{i}=\pm1$\ of the field. For $g\neq0$, there is a single
true equilibrium state (a property that we shall use in section 7), which
corresponds to the absolute minimum of (3.3). It is reached for the solution
with largest $\left\vert m\right\vert $\ of the equations%
\begin{equation}
m_{i}=\tanh\left(  h_{i}/T\right)  \text{\quad,\quad}h_{i}=\pm g+Jm^{3}%
\tag{3.5}\label{003.005}%
\end{equation}
($m_\uparrow>0$, $m_\downarrow<0$). For $g=0$, below the Curie temperature
$T=0.36$\textrm{~}J, the invariance $m\longleftrightarrow-m$\ is spontaneously
broken:  
there are two stable ferromagnetic states, with $m_\uparrow=-m_\downarrow
=m^{\mathrm{f}}$\ very close to $1$; for $T\ll J$, $1-m^{\mathrm{f}}%
\sim2e^{-J/T}$. The paramagnetic state $m^{\mathrm{p}}=0$ is still a local
minimum of (3.3); this is why we chose a quartic rather than a quadratic
interaction. As $g$ increases, this paramagnetic state is shifted, with
$m^{\mathrm{p}}\sim g/T$, but it remains metastable. It disappears 
only when $g$
becomes sufficiently large, in which case the single 
minimum of $F_{\uparrow
}\left(  m_{\uparrow}\right)  $ is the ferromagnetic one; the limiting value
$g_{\mathrm{c}}$ is found as function of $J$ and $T$ by eliminating $m$ from%
\begin{gather}
m=\tanh\left(  g_{\mathrm{c}}/T+Jm^{3}/T\right),\nonumber\\
2m^{2}=1-\sqrt{1-4T/3J}\text{ ,}\tag{3.6}\label{003.006}%
\end{gather}
which yields $m^{2}\sim \frac{T}{3J}$, 
$g_{\mathrm{c}}^{2}\sim \frac{4T^{3}}{27J}$ when $T\ll
J$.

These features agree with what we expect for a measuring apparatus. As
the initial metastable state of $\mathrm{A}$ we can take
$\mathcal{R}\left( 0\right) =R_{\mathrm{M}}\left( 0\right) \otimes
R_{\mathrm{B}}$\ where $R_{\mathrm{M}}\left( 0\right) =1/2^{N}$\ is
the paramagnetic state of the magnet, with $m=\left\langle
\hat{m}\right\rangle =0$, and $R_{\mathrm{B}} $\ is the equilibrium
density operator $R_{\mathrm{B}}\propto e^{-\hat {H}_{\mathrm{B}}/T}$\
of the phonon bath. This state has a lifetime long as an exponential
of $N$ for $N\gg1$\ since (3.3) is a minimum, and can easily be
prepared by cooling $\mathrm{A}$ before the measurement. The
transition towards either one or the other of the two stable
ferromagnetic states $\mathcal{R}_{i}=R_{\mathrm{M}i}\otimes
R_{\mathrm{B}}$ ($i=\uparrow$ or $\downarrow$)\ is expected to be
\textit{triggered} by the interaction (3.2), depending on $r\left(
0\right) $. The macroscopic size of $\mathrm{A}$ allows such a
relaxation to be irreversible, and the stability of the two states
$\mathcal{R}_{i}$\ ensures a robust and {\it permanent
registration}. The fact that, in each state $R_{\mathrm{M}i}$, $m$
has statistical fluctuations around $\pm m^{\mathrm{f}}$ small as
$1/\sqrt{N}$\ ensures a clear distinction between the two possible
outcomes $\pm m^{\mathrm{f}}$ since $m^{\mathrm{f}}$ is close
to $\pm1$. Thus $\left\langle \hat{m}\right\rangle $, which in the
final state may take two values in correspondence with the eigenvalues
of $\hat{s}_{z}$, is a good candidate for a pointer variable. Finally
the fact that $\mathrm{M}$ displays for $N\gg1$ a \textit{phase
transition} with broken symmetry makes the two outcomes equally
probable a priori, and thus prevents bias of the apparatus.

The spin system $\mathrm{M}$, although large, cannot reach equilibrium by
itself since its Hamiltonian is too simple. This will be achieved owing to the
phonon bath $\mathrm{B}$, described by an independent phonon Hamiltonian
$\hat{H}_{\mathrm{B}}$\ with a dense, quasi continuous spectrum and a Debye
frequency cutoff $\Gamma$.\ A weak interaction ($\gamma\ll1$)
\begin{equation}
\hat{H}_{\mathrm{MB}}=\sqrt{\gamma}\sum_{n,a}\hat{\sigma}_{a}^{\left(
n\right)  }\hat{B}_{a}^{\left(  n\right)  }\tag{3.7}\label{003.007}%
\end{equation}
involving phonon operators $\hat{B}_{a}^{\left(  n\right)  }$ ($a=x,y,z$ ;
$n=1,\ldots N$)\ is able to thermalize the spins after some rather long delay
of order $\hbar/\gamma T$. Since $\gamma$ is small and the bath large, the
correlations between $\mathrm{M}$ and $\mathrm{B}$ are negligible in the
initial state $\mathcal{R}\left(  0\right)  $, and the marginal state of
$\mathrm{B}$, initially at equilibrium $R_{\mathrm{B}}\propto e^{-\hat
{H}_{\mathrm{B}}/T}$, is not deeply affected by the evolution.

Altogether the full Hamiltonian of $\mathrm{S}+\mathrm{A}$ is given by
$\hat{H}=\hat{H}_{\mathrm{SA}}+\hat{H}_{\mathrm{M}}+\hat{H}_{\mathrm{B}}
+\hat{H}_{\mathrm{MB}}$. It commutes with the measured observable 
$\hat{s}_{z}$; this is a standard requirement \cite{way},  
which ensures that the measured quantity does not change during the
measurement process. In section 6 we shall replace (3.2) by the more general
spin-apparatus interaction%
\begin{equation}
\hat{H}_{\mathrm{SA}}'=-\sum_{n}g_{n}
\hat{s}_{z}
\hat{\sigma}_{z}^{\left(  n\right)  }%
\text{ ,}\tag{3.8}\label{003.008}%
\end{equation}
for which the coupling constants $g_{n}$\ between the measured spin $\hat
{s}_{z}$\ and the various apparatus spins $\hat{\sigma}_{z}^{\left(  n\right)
}$\ are not the same, due to the different locations of the spins of
$\mathrm{M}$ in the magnetic grain. We assume the deviation 
$\delta g^{2}=N^{-1}\sum_{n}\left(  g_{n}-g\right)  ^{2}$\ of the couplings $g_{n}%
$\ around their average $g$\ to be small, $\delta g\ll g$.

\section{Dynamical equations}

Our purpose is to solve the equation of motion (2.3), starting from the
initial condition $\mathcal{D}\left(  0\right)  =r\left(  0\right)  \otimes
R_{\mathrm{M}}\left(  0\right)  \otimes R_{\mathrm{B}}$. The large value of
$N$\ and the form of $\hat{H}_{\mathrm{M}}$\ and $\hat{H}_{\mathrm{SA}}%
$\ suggest us to rely on a time-dependent mean-field approach, which we expect
to become exact as $N\rightarrow\infty$.  
We only sketch below the main results and the main steps of the derivation;
details, proofs and discussions will be found in \cite{ABNprep}.

We first note that owing to the conservation of $\hat{s}_{z}$, the four blocks
of $\mathcal{D}\left(  t\right)  $\ labeled by the eigenvalues $i=\uparrow$ or
$\downarrow$\ of $\hat{s}_{z}$, initially proportional to $r_{\uparrow
\uparrow}\left(  0\right)  $, $r_{\uparrow\downarrow}\left(  0\right)  $,
$r_{\uparrow\uparrow}\left(  0\right)  $, $r_{\downarrow\downarrow}\left(
0\right)  $, evolve independently. The equations of motion in each sector have
in the Hilbert space of $\mathrm{A}$ the form%
\begin{equation}
i\hbar\frac{d\mathcal{D}_{ij}}{dt}=-gN\left(  s_{i}\hat{m}\mathcal{D}%
_{ij}-\mathcal{D}_{ij}s_{j}\hat{m}\right)  +\left[  \hat{H}_{\mathrm{A}%
},\mathcal{D}_{ij}\right]  \text{ ,}\tag{4.1}\label{004.001}%
\end{equation}
where $\hat{H}_{A}=\hat{H}_{\mathrm{M}}+\hat{H}_{\mathrm{B}}+\hat
{H}_{\mathrm{MB}}$. Our mean-field approximation consists in replacing at each
time $\hat{H}_{\mathrm{M}}$\ in each of the four equations (4.1) by
$m_{ij}^{4}+4m_{ij}^{3}\left(  \hat{m}-m_{ij}\right)  $, and using the
self-consistency condition%
\begin{equation}
m_{ij}=\frac{\operatorname{Tr}_{\mathrm{A}}\hat{m}\left\vert \mathcal{D}%
_{ij}\right\vert }{\operatorname{Tr}_{\mathrm{A}}\left\vert \mathcal{D}%
_{ij}\right\vert }\text{\quad,\quad}\left\vert \mathcal{D}_{ij}\right\vert
\equiv\sqrt{\mathcal{D}_{ij}\mathcal{D}_{ij}^{\dag}}\text{ .}\tag{4.2}%
\label{004.002}%
\end{equation}
This approach differs from standard mean-field approaches through the
occurrence of \textit{different parameters} $m_{ij}$\ in the four sectors.
They have only the status of mathematical tools, and their simultaneous
occurrence in the equations of motion shows that they are not directly related
to the expectation value of $\hat{m}$\ in the state $\mathcal{D}\left(
t\right)  $, contrary to what happens in usual mean-field theories such as
(3.5). Moreover the unconventional form of (4.2) is related to the
non-hermiticity of $\mathcal{D}_{ij}=\mathcal{D}_{ji}^{\dag}$. The present
approach can be justified by showing that the corrections are negligible in
the large $N$\ limit.

The phonon variables are eliminated through replacement of (4.1) by an
equation for the partial trace $D_{ij}\left(  t\right)  =\operatorname{Tr}%
_{\mathrm{B}}\mathcal{D}_{ij}\left(  t\right)  $, an operator in the Hilbert
space of $\mathrm{M}$. This is achieved, as usual in the limit of a very large
bath $\mathrm{B}$ weakly coupled ($\gamma\ll1$)\ with $\mathrm{M}$,  
by noting that the bath occurs only through the \textit{memory kernel}
$\operatorname{Tr}_{\mathrm{B}}R_{\mathrm{B}}\hat{B}_{a}^{\left(  n\right)
}\left(  t\right)  \hat{B}_{b}^{\left(m\right)}
\left(  t^{\prime}\right)  $. For simplicity
we choose the bath Hamiltonian $\hat{H}_{\mathrm{B}}$\ in such a way that this
kernel has the form $\delta_{a,b}\delta_{n,m}K\left(  t-t^{\prime}\right)  $,
with a quasi-ohmic spectrum:%
\begin{equation}
K\left(  t\right)  =\hbar^{2}\int_{-\infty}^{+\infty}\frac{d\omega}{16\pi
}e^{i\omega t}\omega\left[  \coth\left(  \hbar\omega/2T\right)  -1\right]
e^{-\left\vert \omega\right\vert/ \Gamma}\text{ .}\tag{4.3}\label{004.003}%
\end{equation}
The memory time $\hbar/T$\ of $\mathrm{B}$ is taken much smaller than the
equilibration time $\hbar/\gamma T$\ of $\mathrm{M}$.

According to the mean-field surmise, the operator $D_{ij}$\ for each pair $ij
$\ can be \textit{factorized} as%
\begin{equation}
D_{ij}\left(  t\right)  =r_{ij}\left(  0\right)  \times\rho_{ij}^{\left(
1\right)  }\left(  t\right)  \otimes\cdots\otimes\rho_{ij}^{N}\left(
t\right)  \text{ ,}\tag{4.4}\label{004.004}%
\end{equation}
where each factor%
\begin{equation}
\rho_{ij}^{\left(  n\right)  }=\frac{1}{2}\left(  \zeta_{0,ij}+\sum_{a}%
\zeta_{a,ij}\hat{\sigma}_{a}^{\left(  n\right)  }\right)  =\rho_{ji}^{\left(
n\right)  \dag}\tag{4.5}\label{004.005}%
\end{equation}
is an operator in the $2$-dimensional Hilbert space of the spin $n$\ of
$\mathrm{M}$, parametrized by the 4 functions $\zeta_{0,ij}$, $\zeta_{a,ij}$
($a=x,y,z$)\ of time. The elimination of the bath variables then provides for
these functions equations of motion which depend on $K\left(  t\right)  $\ in
a rather complicated way. We write them here in a simplified form, which is
valid for the time scales over which they will be relevant, namely%
\begin{gather}
\dot{\zeta}_{0,\uparrow\downarrow}=\frac{2ig}{\hbar}\zeta_{z,\uparrow
\downarrow},\nonumber\\
\dot{\zeta}_{z,\uparrow\downarrow}=\frac
{2ig}{\hbar}\left(  1+\frac{\gamma\Gamma^{2}t^{2}}{2\pi}\right)
\zeta_{0,\uparrow\downarrow}-\frac{\gamma\Gamma^{2}t}{\pi}\zeta_{z,\uparrow
\downarrow},\tag{4.6}\label{004.006}%
\end{gather}
for short times ($t\ll1/\Gamma$), and%
\begin{equation}
\dot{\zeta}_{0,ii}=0\text{\quad,\quad}\dot{m}_{i}=\frac{\gamma h_{i}}{\hbar
}\left(  1-\frac{m_{i}}{\tanh h_{i}/T}\right) \tag{4.7}\label{004.007}%
\end{equation}
for larger times ($t\gg\hbar/T$). The parameters $m_{i}\equiv\zeta
_{z,ii}/\zeta_{0,ii}$\ introduced in (4.7) are the time-dependent
magnetizations in the two diagonal sectors and we denote here as $h_{i}%
=gs_{i}+Jm_{i}^{3}$\ with $s_{i}=\pm1$\ the associated effective fields. The
parameters $m_{\uparrow\downarrow}$\ and $m_{\downarrow\uparrow}$\ are found
to vanish, as well as $\zeta_{x,ij}$\ and $\zeta_{y,ij}$. The initial
paramagnetic conditions are $\zeta_{0,ij}=1$, $\zeta_{a,ij}=0$.
It can also be shown that in the limit of a large bath B and of a
weak coupling $\gamma$, the state of B is hardly affected and that the
correlations between B and M are negligible.

We stress that, although our equations (4.6) and (4.7) have been
obtained through approximations, they become exact in the limit of a
large apparatus. On the one hand, the large value on $N$\ together
with the long range of the interaction $\hat{H}_{\mathrm{M}} $\
ensure the validity of the mean-field approach and the existence of
different phases for $\mathrm{M}$. On the other hand, the
quasi-continuity of the phonon spectrum entails that the energy
exchanges between $\mathrm{M}$ and $\mathrm{B}$ dissipate entropy on
any reasonable time lapse.

\section{Initial collapse of $\mathrm{S}$}

Let us first solve eq. (4.6) for
$\mathcal{D}_{\uparrow\downarrow}\left( t\right)$ and for
$\mathcal{D}_{\downarrow\uparrow}\left( t\right) 
=\mathcal{D}_{\uparrow\downarrow}\left( t\right)^{\dagger}$. Over very short
times ($t<\hbar/g$), the evolution is governed only by the interaction
$\hat{H}_{\mathrm{SA}}$, and we have%
\begin{equation}
\zeta_{0,\uparrow\downarrow}=\cos\frac{2gt}{\hbar}\text{\quad,\quad}%
\zeta_{z,\uparrow\downarrow}=i\sin\frac{2gt}{\hbar}\tag{5.1}\label{005.001}.%
\end{equation}
Hence, the off-diagonal block $\mathcal{D}_{\uparrow\downarrow}^{\left(
t\right)  }$\ of $\mathcal{D}\left(  t\right)  $\ behaves on this time scale
as%
\begin{equation}
\mathcal{D}_{\uparrow\downarrow}\left(  t\right)  
=\frac{r_{\uparrow\downarrow}\left(
0\right)  }{2^{N}}D_{\mathrm{B}}\otimes%
%TCIMACRO{\dprod \limits_{n=1}^{N}}%
%BeginExpansion
{\displaystyle\prod\limits_{n=1}^{N}}
%EndExpansion
\left(  \cos\frac{2gt}{\hbar}+i\sin\frac{2gt}{\hbar}\hat{\sigma}_{z}^{\left(
n\right)  }\right)  \text{ .}\tag{5.2}\label{005.002}%
\end{equation}

As a consequence, the element $r_{\uparrow\downarrow}\left(  t\right)  $\ of
the marginal density operator of $\mathrm{S}$\ rapidly decreases as%
\begin{equation}
r_{\uparrow\downarrow}\left(  t\right)  =r_{\uparrow\downarrow}\left(
0\right)  \cos^{N}\frac{2gt}{\hbar}\sim r_{\uparrow\downarrow}\left(
0\right)  e^{-t^{2}/\tau_{\operatorname{red}}^{2}},
\tag{5.3}\label{005.003}%
\end{equation}
over a \textit{reduction time}%
\begin{equation}
\tau_{\operatorname{red}}=\frac{1}{\sqrt{2N}}\frac{\hbar}{g}\text{ ,}%
\tag{5.4}\label{005.004}%
\end{equation}
which will be the shortest of all the characteristic times that we shall
encounter in the dynamics of $\mathrm{S}+\mathrm{A}$. This decay, 
which describes a very rapid collapse of the off-diagonal terms of 
$r(t)$, leads to
negligible values of (5.3), small as $e^{-N}$\ for $t=\hbar/g\sqrt{2}$.

From the viewpoint of the system $\mathrm{S}$, such a behaviour is reminiscent
of a decoherence. However, instead of 
the factor
$g$ occurring in the denominator of the reduction time
(5.4), environment induced decoherence times involve 
the temperature of the environment. More crucially, whereas a usual
decoherence takes place in a given basis, the \textit{basis} in which the
off-diagonal components disappear is \textit{determined here by the
observable} which is being measured. A rotation of the apparatus, which is
anisotropic, modifies this basis in a controlled way.

Eq.(5.2) shows that the disappearance of the off-diagonal elements of
$r\left(  t\right)  $, that is, of $\left\langle \hat{s}_{x}\right\rangle
$\ and $\left\langle \hat{s}_{y}\right\rangle $, is accompanied by the
\textit{creation of correlations} between $\hat{s}_{x}$\ (or $\hat{s}_{y}$)
and an arbitrary number of $\hat{\sigma}_{z}^{\left(  n\right)  }$\ operators.
As usual in relaxation processes, the relaxation of $\left\langle \hat{s}%
_{x}\right\rangle $\ and $\left\langle \hat{s}_{y}\right\rangle $\ is
compensated for by a transfer of order towards more and more
complicated observables coupling $\mathrm{S}$ and $\mathrm{M}$. 
The number of these correlations is 
extremely large, but each one is
{\it very small}; indeed, if they involve a number $p\ll N$ of
spins of the apparatus, their value 
$r_{\uparrow\downarrow}\left(
0\right)  e^{-t^{2}/\tau_{\operatorname{red}}^{2}}\left(  2gt/\hbar\right)
^{p}$ is hindered
either due to the exponential or
due to the small value of $t$ in the range $\tau_{\rm red}$.
This is an example of transfer of order towards
many-spin correlations (of order $N$) that we alluded to when recalling the
mechanism of irreversibility.

\section{Suppression of recurrences}

The expression (5.2) for $\mathcal{D}_{\uparrow\downarrow}\left(  t\right)  $,
valid for the very beginning $t<\hbar/g$\ of the measurement process, would
exhibit a periodic structure, with period $t=\pi\hbar/g$, if we
extrapolated it
towards larger times. According to (5.3), $r_{\uparrow\downarrow}\left(
t\right)  $, which nearly vanishes after a time 
$\tau_{\operatorname{red}}$,
would present later on, for integer values of $2gt/\pi\hbar$, a sequence of
narrow gaussian peaks, all with height $r_{\uparrow\downarrow}\left(  0\right)
$. The collapse would then not be irreversible.

However, when establishing (5.1), we have dropped the bath terms of (4.6),
which must be taken into account as $t$\ increases. Their effect, apart from
slightly shifting the position of the peaks, is to multiply (5.1) by
$e^{-\chi\left(  t\right)  }$, with $\chi\left(  t\right)  \sim\gamma
\Gamma^{2}g^{2}t^{4}/2\pi\hbar^{2}$. Hence, the off-diagonal parts of not only
the marginal density operator of $\mathrm{S}$ but also of the full state of
$\mathrm{S}+\mathrm{A}$, that is, 
both (5.3) and (5.2) are damped by the factor
\begin{equation}
e^{-t^{4}/\tau_{\mathrm{2}}^{4}}\text{\quad,\quad}\tau_{\mathrm{2}%
}=\left(  \frac{2\pi}{\gamma N}\right)  ^{1/4}\left(  \frac{\hbar}{\Gamma
g}\right)  ^{1/2}\text{ .}\tag{6.1}\label{006.001}%
\end{equation}
This \textit{decay time} $\tau_2$, 
which has the same nature as the off-diagonal spin-lattice relaxation
time in NMR, is much larger than the
collapse time (5.4) if $\gamma\ll 1$
and $N\gg 1$. We choose the parameters of the model in
such a way that it is much smaller than the first recurrence time $\pi
\hbar/2g$, that is, $\gamma\gg g^{2} /N\hbar^{2}\Gamma^{2}$. All recurrent
peaks are therefore cancelled; the height of the first one in 
$r_{\uparrow\downarrow
}\left(  t\right)  $\ is small as $r_{\uparrow\downarrow}\left(  0\right)
e^{-N\pi^{3}\gamma\hbar^{2}\Gamma^{2}/32g^{2}}$\ and the \textit{full matrix}
$\mathcal{D}_{\uparrow\downarrow}\left(  t\right)  $\ disappears on the time
scale $\tau_2$.

The presence of the phonon bath thus \textit{makes the collapse irreversible}.
During the first stage, over the characteristic 
time $\tau_{\rm red}$, we have seen that the
off-diagonal order initially present in $\mathrm{S}$, 
which is expressed by a finite value of $\left\langle
\hat{s}_{x}\right\rangle $\ or $\left\langle \hat{s}_{y}\right\rangle $,
dissolves into correlations between $\mathrm{S}$ and $\mathrm{M}$, but this
transfer is reversible as shown by the possibility of recurrences. During the
next stage, over the time scale $\tau_2$, 
the order is transferred further to the phonon bath,
but now can no longer come back after any reasonable time lapse.

In fact, the recurrent peaks in $r_{\uparrow\downarrow}\left(  t\right)
$\ may disappear even if there is no bath, provided the interaction 
between S and A has the form 
$\hat{H}_{\mathrm{SA}}'$\
given by (3.8). The solution for $\gamma=0$\ of the
equations of motion for $\mathcal{D}_{\uparrow\downarrow}\left(  t\right)
$\ is then given at all times by
\begin{gather}
\tag{6.2}\label{006.002}
\mathcal{D}_{\uparrow\downarrow}
\left(  t\right)  =\frac{r_{\uparrow\downarrow}\left(
0\right)  }{2^{N}}D_{\mathrm{B}}%\\
\otimes
{\displaystyle\prod\limits_{n=1}^{N}}
\left(  \cos\frac{2g_{n}t}{\hbar}+i\sin\frac{2g_{n}t}{\hbar}\hat{\sigma}%
_{z}^{\left(  n\right)  }\right)  \text{ .}
%%\nonumber
\end{gather}
Instead of (5.3) we find a destructive interference of the $\cos$ factors
entering%
\begin{gather}
r_{\uparrow\downarrow}\left(  t\right)  =r_{\uparrow\downarrow}\left(
0\right)
{\displaystyle\prod\limits_{n=1}^{N}}
\cos\frac{2g_{n}t}{\hbar}%\nonumber\\
\sim r_{\uparrow\downarrow}\left(  0\right)
e^{-t^{2}/\tau_{\mathrm{2}}^{\prime^{2}}}\cos^{N}\frac{2gt}{\hbar}\text{
,}\tag{6.3}\label{006.003}%
\end{gather}
which again produces a decay, with the alternative characteristic time%
\begin{equation}
\tau_{\mathrm{2}}^{\prime}=\frac{1}{\sqrt{2N}}\frac{\hbar}{\delta g}\text{
.}\tag{6.4}\label{006.004}%
\end{equation}
The height of the first peak is here small as $r_{\uparrow\downarrow}\left(
0\right)  e^{-N\pi^{2}\delta g^{2}/2g^{2}}$. A dispersion such that
$1\gg\delta g/g\gg1\sqrt{N}$ is therefore sufficient to make all the
recurrent peaks disappear. Provided $N$ is large and the couplings are
slightly different, the \textit{bath is not necessary} to ensure the
irreversible disappearance of the off-diagonal elements $r_{\uparrow
\downarrow}\left(  t\right)  $.

The existence of two alternative mechanisms which  
make the rapid disappearance (5.3) of the off-diagonal parts of $r\left(
t\right)  $\ irreversible is reminiscent of two alternative relaxation
mechanisms in NMR. The decay (6.1) due to the coupling with the phonons
looks like a spin-lattice relaxation, while the decay (6.2) due to a
dispersion in the couplings $g_{n}$\ looks like the relaxation due to the
spread of the Larmor frequencies in the transverse motion of non interacting
spins in an inhomogeneous magnetic field. As in the latter case, the initial
order associated with the non vanishing value of $\left\langle \hat{s}%
_{x}\right\rangle $\ or $\left\langle \hat{s}_{y}\right\rangle $, which has
escaped towards correlations, gets trapped there due to the inhomogeneity of
the couplings $g_{n}$\ as exhibited by (6.2). However, we may imagine to
retrieve this order, as currently done in NMR by means of spin-echo
experiments. Suppose it is possible to apply on $\mathrm{M}$\ an external
magnetic field which acts on the spins $\overrightarrow{\hat{\sigma}}^{\left(
n\right)  }$\ without affecting $\mathrm{S}$. Then a brief pulse $\pi$\ around
$y$\ applied at the time $\theta$\ suddenly changes the signs of the operators
$\hat{\sigma}_{z}^{\left(  n\right)  }$\ in the expression (6.2) for the state
$\mathcal{D}_{\uparrow\downarrow}\left(  \theta\right)  $. The subsequent
evolution generated by $\hat{H}_{\mathrm{SA}}'$ will lead to $\mathcal{D}%
_{\uparrow\downarrow}\left(  2\theta\right)  =\mathcal{D}_{\uparrow\downarrow
}\left(  0\right)  $\ and hence to $r_{\uparrow\downarrow}\left(
2\theta\right)  =r_{\uparrow\downarrow}\left(  0\right)  $. The second
relaxation mechanism is therefore less effective than the first one. 

Anyhow,
even though we do not need a phonon bath to explain why the final state (1.2)
of $\mathrm{S}+\mathrm{A}$ involves no off-diagonal block, we must resort to
it to account for the final form of the diagonal blocks, which results
from energy exchanges between M and B.

\section{Registration}

Let us now turn to the evolution of $\mathcal{D}_{ii}\left(  t\right)  $\ for
$i=\uparrow$ and $i=\downarrow$, which is governed by (4.7). 

We first have at all time $\zeta_{0,ii}\left(  t\right)  =1$ 
and hence $\zeta
_{z,ii}=m_{i}$. We can rewrite the equation for $m_{\uparrow}\left(  t\right)
$, with $m_{\uparrow}\left(  0\right)  =0$, as%
\begin{gather}
\frac{\hbar}{\gamma}\frac{dm_{\uparrow}}{dt}=h_{\uparrow}\left(
1-\frac{m_{\uparrow}}{\tanh h_{\uparrow}/T}\right)  
%%R\nonumber\\
=-\frac{dF_{\uparrow}%
}{dm_{\uparrow}}\frac{1-\frac{m_{\uparrow}}{\tanh h_{\uparrow}/T}}%
{1-\frac{\tanh^{-1}m_{\uparrow}}{h_{\uparrow}/T}}\text{ ,}\tag{7.1}%
\label{007.001}%
\end{gather}
where $h_{\uparrow}\equiv g+Jm_{\uparrow}^{3}$\ and where the function
$F_{\uparrow}\left(  m_{\uparrow}\right)  $\ is defined by (3.3); for
$m_{\downarrow}\left(  t\right)  $, $g$\ will be changed into $-g$. The last
factor in (7.1) is a positive function of $m_{\uparrow}$. Hence, $m_{\uparrow
}\left(  t\right)  $\ relaxes by increasing up to the smallest positive value
of $m$\ such that $F_{\uparrow}\left(  m\right)  $\ is minimal. As discussed
in section 3, this minimum is the paramagnetic one, not only for $T$\ above the
transition temperature, but also below if $g$\ is smaller than $g_{\mathrm{c}%
}$\ given by (3.6). In such a case, the measurement fails since suppression of
the interaction term $\hat{H}_{\mathrm{SA}}$\ would bring back the apparatus
to its initial state with $m_{\uparrow}=0$.

Let us therefore restrict to $g>g_{\mathrm{c}}$. Numerically, if we take
$T=0.34J$, slightly below the transition temperature $0.36J$, we have
$g_{\mathrm{c}}=0.08J$. In this
 case, $m_{\uparrow}\left(  t\right)  $\ increases up to the value
$m^{\mathrm{f}}$\ where $F_{\uparrow}\left(  m\right)  $\ has its lowest
minimum, so that $\mathrm{M}$ reaches the ferromagnetic state with $m$\ very
close to $+1$\ ($m^{\mathrm{f}}=0.996J$\ for $T=0.34J$\ and $g=0.09J$). If the
coupling between $\mathrm{S}$ and $\mathrm{A}$ is then switched off,
$m_{\uparrow}$\ remains practically unchanged near $m^{\mathrm{f}}$.

In the sector $\mathcal{D}_{\downarrow\downarrow}$\ of the density
matrix, $\mathrm{M}$ symmetrically reaches the ferromagnetic state
with $m_{\downarrow }\left( t\right) $\ tending to
$-m^{\mathrm{f}}$. The memory of the triggering of $\mathrm{A}$ by
$\mathrm{S}$ is kept forever. The overall density matrix
$\mathcal{D}\left( t\right) $\ therefore reaches the expected form
(1.2) and all the features of ideal measurements listed above are
obtained.

The time dependence of $m_{\uparrow}\left(  t\right)  $\ or $m_{\downarrow
}\left(  t\right)  $\ is found from (7.1) by integration. Contrary to the
characteristic time scales (5.4), (6.1) and (6.4) for $\mathcal{D}%
_{\uparrow\downarrow}$\ and $\mathcal{D}_{\downarrow\uparrow}$, the time
scales for $\mathcal{D}_{\uparrow\uparrow}$\ and $D_{\downarrow\downarrow}%
$\ are not divided by a power of $N$\ and are thus much longer. 
We illustrate the behaviour as function of $m_{\uparrow}$\ of the right-hand
side of (7.1) by considering the regime $g\ll T\ll J$. Equal to $g$\ for
$m_{\uparrow}=0$, 
this right-side first decreases down to $g-g_{\mathrm{c}}$,
$g_{\mathrm{c}}^{2}=4T^{3}/27J$, a minimum reached for $m_{\uparrow}^{2}%
=T/3J$. It then increases up to the value $27J/256$\ attained for
$m_{\uparrow}=3/4$, and decreases again to vanish for $m_{\uparrow
}=m^{\mathrm{f}}$\ as $J\left(  m^{\mathrm{f}}-m_{\downarrow}\right)  $. Hence
$m_{\uparrow}\left(  t\right)  $\ approaches $m^{\mathrm{f}} $\ for large
$t$\ asymptotically as $e^{-\gamma Jt/\hbar}$. Although strictly speaking the
full relaxation time is thus infinite, $m_{\uparrow}\left(  t\right)
$\ reaches values nearly equal to $m^{\mathrm{f}}$\ (with $m^{\mathrm{f}}%
-m$\ of order $g/J$) after a finite delay $\tau_{\mathrm{reg}} $\ governed by
the region $m^{3}\ll T/J$. Integration of (7.1) then provides
\begin{gather}
\frac{\gamma t}{\hbar}=\int_{0}^{m_{\uparrow}\left(  t\right)  }\frac
{dm}{g+Jm^{3}-Tm} \nonumber\\
=\frac{3}{T}\int_{0}^{m_{\uparrow}\left(  t\right)
\sqrt{3J/T}}\frac{dx}{\left(  x-1\right)  ^{2}\left(  x+2\right)  +2\left(
g-g_{\mathrm{c}}\right)  /g_{\mathrm{c}}}.
\tag{7.2}
\label{007.002}
\end{gather}
The \textit{registration time} thus found
\begin{gather}
\tau_{\mathrm{reg}}=\frac{3\hbar}{\gamma T}\int_{0}^{\infty}\frac{dx}{\left(
x-1\right)  ^{2}\left(  x+2\right)  +2\left(  g-g_{\mathrm{c}}\right)
/g_{\mathrm{c}}},\nonumber\\
g_{\mathrm{c}}=\frac{2T}{3}\sqrt{\frac
{T}{3J}},\tag{7.3}\label{007.003}%
\end{gather}
behaves as%
\begin{equation}
\tau_{\mathrm{reg}}=\frac{\pi\hbar}{\gamma T}\sqrt{\frac{3g_{\mathrm{c}}%
}{2\left(  g-g_{\mathrm{c}}\right)  }}\tag{7.4}\label{007.004}%
\end{equation}
for $g-g_{\mathrm{c}}\ll g_{\mathrm{c}}$. 
For $g-g_{\mathrm{c}}$ of order $g_{\mathrm{c}}$,  
$\tau_{\mathrm{reg}}$ is proportional to $\hbar/\gamma T$ and 
thus depends
only on the bath; it becomes large for a weak coupling $\gamma$ between 
M and B and at low bath temperature.

\section{Conclusion}

In spite of its simplicity, the model gives rise to an elaborate scenario. It
produces all the required features of a quantum measurement, and exhibits
several time scales. During the very first stages, the large apparatus,
without changing much, destroys the off-diagonal elements of the density
matrix of $\mathrm{S}$. This takes place very rapidly, over the reduction time
(5.4), and the possible recurrences are hindered owing to interaction with the
phonon bath. In spite of the weakness of this interaction, the corresponding
decay time (6.1) can be short. The inertia of the large apparatus implies that
its changes are much slower; they occur significantly only 
through correlations
with the diagonal elements of the state of $\mathrm{S}$. They are triggered by
the coupling $g$\ between $\mathrm{S}$ and $\mathrm{A}$, and take
place on the registration
time scale (7.3) or (7.4) independent of the size of $\mathrm{M}%
$. After this time, the coupling $g$\ is ineffective.

The dynamical process that we described shows that the experiment can 
behave as a
measurement only if the parameters satisfy the inequalities%
\begin{equation}
N\gg1\text{\quad,\quad}N\gg\frac{1}{\gamma}\left(  \frac{g}{\hbar\Gamma
}\right)  ^{2}\text{\quad or\quad}N\gg\frac{g^{2}}{\delta g^{2}},%
\tag{8.1}\label{008.001}%
\end{equation}
which ensure the disappearance of the off-diagonal blocks, and
\begin{equation}
\hbar\Gamma\gg T\gg\gamma J\text{\quad,\quad}\hbar\Gamma\gg J>g\text{
,}\tag{8.2}\label{008.002}%
\end{equation}
which were used to establish the relaxation equations.

The above solution enforces the statistical interpretation of quantum
mechanics, according to which a density operator (even when it reduces
to the projection on a pure state) plays with respect to the
non-commuting observables the same 
${\rm r\hat{o}le}$ as the probability density
with respect to the commuting physical variables in classical
statistical mechanics. In this interpretation, a quantum
\textquotedblleft state\textquotedblright\ does not refer to a single object
but only characterizes the statistics of an ensemble of identically prepared
objects. Only a probabilistic description of the microscopic world is
available to us. 
Moreover, since a quantum measurement requires a macroscopic apparatus,
it can be described only by means of quantum statistical physics. The
irreversible aspects of quantum measurements are then explained by the large
size of the apparatus, as in the solution of the irreversibility paradox. Here
too, we need in practice
to rely on approximation schemes but our approach becomes exact
in the limit of a large apparatus over any reasonable delay; the consideration
of time scales is crucial.

A specific feature arises from the dynamics of the measurement
process. The laws of quantum mechanics involve a special type of
probabilities associated with the non commutation of the observables,
as exemplified by Bell's inequalities. These laws even violate
standard logical reasoning. In the GHZ paradox \cite{GHZ} three statements
separately true, but which can be checked experimentally only by means
of different measurement settings, are not true together: if they are
put together, they have a common consequence which can be shown
experimentally to be wrong. The above solution of a model for
measurement shows that quantum mechanics is consistent, although its
assertions are {\it contextual}: they are valid only in a given
experimental setting governed by the measuring apparatus. We have
shown how the process generates a final state of the form (1.2), which
does not involve off-diagonal blocks. Their disappearance, which is a
real dynamical phenomenon, is conceptually important since they have
no classical meaning, nor even ordinary logical interpretation. Thus,
\textit{classical probabilities emerge} from an initial state $r\left(
0\right)  $\ of $\mathrm{S}$ which cannot be described classically. This
possibility of interpreting the outcome of 
a measurement in a classical language arises
owing to the \textit{change of scale}, just as continuity of matter 
or phase transitions or
irreversibility emerge from a large number of degrees of freedom. Moreover, a
change of orientation of the apparatus allows us to explore other components
of $r\left(  0\right)  $, in agreement with the contextual nature of quantum mechanics.

We have also seen how \textit{classical correlations} emerge in the
diagonal blocks of (1.2) for an ideal measurement. These correlations
allow us to get information on $\mathrm{S}$ through registration in
$\mathrm{M}$. The non standard features of quantum correlations have
been lost together with the off-diagonal terms.  This loss of
off-diagonal information is the price to be paid for gaining classical
information about the diagonal elements of $r\left( 0\right) $.
Let us stress that the Born rule and the von Neumann reduction have
been recovered in our approach because we could 
interpret the outcomes of the
quantum pointer variable as a classical distribution of a macroscopic
classical random variable, correlated with S.

Finally, an ideal measurement followed by a selection of one among the
possible outcomes is identified as a \textit{preparation} of $\mathrm{S}$. A
subensemble with properties controlled by the measurement apparatus is thus
extracted from the whole statistical ensemble. Here again it is the
macroscopic size of the apparatus which allows us to distinguish a selected
value of the pointer variable; we can 
thus rely on its classical correlation with
the system $\mathrm{S}$ so as to set it into a controlled new state.

\end{document}